\documentclass[conference,10pt]{IEEEtran}
\IEEEoverridecommandlockouts

\usepackage{xcolor}
\usepackage{balance}
\usepackage{algorithm}
\usepackage{amsmath, amsfonts, amssymb}  
\usepackage{algorithmicx}
\usepackage{algpseudocode}
\usepackage{cite}
\usepackage{graphicx}
\usepackage{textcomp}
\usepackage{acro}
\usepackage{tikz} 
\usepackage[utf8]{inputenc}
\usepackage{pgfplots} 
\usepackage{pgfgantt}
\usepackage{pdflscape}
\usepackage{changes}
\usepackage{comment}
\usepackage{subfigure}
\usepackage{mathtools}
\usepackage{balance}
\usepackage{pgfplots}
\usepgfplotslibrary{groupplots} 
\usepackage{subcaption}
\usepackage{xcolor}
  \definecolor{Cyan}{RGB}{0,255,255}
  \definecolor{BrightGreen}{RGB}{102,204,0}
\usepackage[font=footnotesize]{caption}
\usepackage{amsthm}
\pgfplotsset{compat=newest}
\usetikzlibrary{plotmarks}
\usetikzlibrary{arrows.meta}
\usepgfplotslibrary{patchplots}
\usepackage{grffile}
\allowdisplaybreaks

\newtheorem{rem}{Remark}
\newtheorem{example}{Example}

\DeclareAcronym{PT}{short=PT, long=potential target}
\DeclareAcronym{ToA}{short=ToA, long=time of arrival}
\DeclareAcronym{AoA}{short=AoA, long=angle of arrival}
\DeclareAcronym{JPDA}{short=JPDA, long=joint probabilistic data association}
\DeclareAcronym{BS}{short=BS, long=base station}
\DeclareAcronym{BP}{short=BP, long=belief propagation}
\DeclareAcronym{pdf}{short=pdf, long=probability density function}
\DeclareAcronym{FoV}{short=FoV, long=field of view}
\DeclareAcronym{RFS}{short=RFS, long=random finite set}
\DeclareAcronym{PPP}{short=PPP, long=Poisson point process}
\DeclareAcronym{MBM}{short=MBM, long=Multi-Bernoulli Mixture}
\DeclareAcronym{EKF}{short=EKF, long=extended Kalman filter}
\DeclareAcronym{GOSPA}{short=GOSPA, long=generalized optimal subpattern assignment}
\DeclareAcronym{ISAC}{short=ISAC, long=integrated sensing and communications}
\DeclareAcronym{DISAC}{short=DISAC, long=distributed integrated sensing and communications}
\DeclareAcronym{MTT}{short=MTT, long=multi-target tracking}
\DeclareAcronym{FISST}{short=FISST, long=finite-set statistics}
\DeclareAcronym{PMBM}{short=PMBM, long=Poisson multi-Bernoulli mixture}
\DeclareAcronym{MHT}{short=MHT, long=multiple-hypothesis tracking}
\DeclareAcronym{VSC}{short=VSC, long=virtual sensing cell}

\begin{document}

\bstctlcite{IEEEexample:BSTcontrol}

\title{Belief Propagation-based Target Handover in Distributed Integrated Sensing and Communication}

\author{
    \IEEEauthorblockN{Liping Bai, Yu Ge, Henk Wymeersch}
    \IEEEauthorblockA{Chalmers University of Technology, Gothenburg, Sweden}
    \thanks{This work has been supported by the Wallenberg Foundations through the Wallenberg AI, Autonomous Systems and Software Program. This work has been supported by the SNS JU project 6G-DISAC under the EU's Horizon Europe research and innovation Program under Grant Agreement No 101139130.}
    \vspace{-10mm}
}

\maketitle

\begin{abstract}
Distributed integrated sensing and communication (DISAC) systems are key enablers for 6G networks, offering the capability to jointly track multiple targets using spatially distributed base stations (BSs). A fundamental challenge in DISAC is the seamless and efficient handover of target tracks between BSs with partially overlapping fields of view, especially in dense and dynamic environments. In this paper, we propose a novel target handover framework based on belief propagation (BP) for multi-target tracking in DISAC systems. By representing the probabilistic data association and tracking problem through a factor graph, the proposed method enables efficient marginal inference with reduced computational complexity. Our framework introduces a principled handover criterion and message-passing strategy that minimizes inter-BS communication while maintaining tracking continuity and accuracy. We demonstrate that the proposed handover procedure achieves performance comparable to centralized processing, yet significantly reduces data exchange and processing overhead. Extensive simulations validate the robustness of the approach in urban tracking scenarios with closely spaced targets. 
\end{abstract}

\begin{IEEEkeywords}
6G, DISAC, tracking, trajectory, target handover, belief propagation.
\end{IEEEkeywords}
\section{Introduction}

The \ac{ISAC} paradigm leverages existing communication infrastructure for sensing and localization, either as a standalone service or as an auxiliary function to support communication~\cite{ISAC}. In urban environments, scalability challenges in communication systems emerge from the high density of users. A common remedy is to partition the service area into cells~\cite{taufique2017planning}, each managed by a single \ac{BS}, with users handed over from one cell to another~\cite{fantacci2000performance}. Extending this divide-and-conquer strategy to \ac{ISAC} results in the concept of \ac{DISAC}~\cite{strinati2024distributedintelligentintegratedsensing,DISAC}.

The \ac{DISAC} setup leads naturally to the multi-sensor \ac{MTT} problem, which involves tracking multiple targets using multiple homogeneous sensors (\acp{BS}, in the context of \ac{DISAC}), each with limited \ac{FoV}~\cite{sandell2008distributed}. \ac{MTT} jointly estimates the number and states of a time-varying set of targets from noisy measurements, while addressing measurement origin uncertainty, missed detections, false alarms, and an unknown target count. Centralized processing of multi-sensor \ac{MTT} can significantly outperform single-sensor approaches~\cite{gostar2020centralized, bar1995multitarget,vo_multi-sensor_2017, scalable-method-group}, but distributed multi-sensor \ac{MTT} without a fusion center introduces further challenges~\cite{van2021distributed}. \textit{Target handover}, where tracking responsibility shifts from one sensor to another as targets move across \acp{FoV}, is an effective strategy to address these challenges~\cite{kim2019optimal}.

\begin{figure}
\centering
{\includegraphics[width=0.92\linewidth]{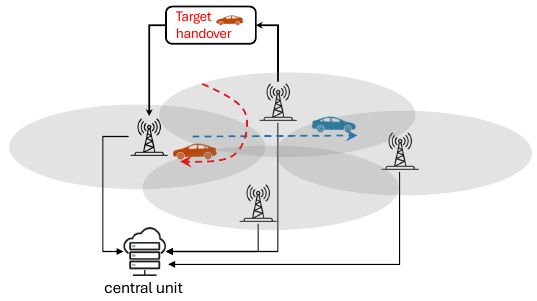}}
\caption{Example of a \ac{DISAC} scenario with a central unit. Four \acp{BS} are shown, with finite and overlapping \acp{FoV}. Target handover enables efficient tracking without needing to share raw data with the central unit.}
\label{fig:overview}
\end{figure}

Recent \ac{DISAC} research has produced various target handover methods. A generalized framework for target mobility management, including dynamic sensing configuration handover, was presented in~\cite{ribeiro_mobility_2025}. A primary \ac{BS} handover within a \ac{VSC}—defined as three neighbouring cells—as well as a handover mechanism between \acp{VSC} was proposed in~\cite{shi}. In~\cite{ge2024target}, a \ac{RFS}-based handover method has been developed. It achieves excellent performance but has poor scalability due to the need to maintain numerous global association hypotheses. In the \ac{MTT} context, \ac{RFS}-based methods~\cite{garcia2018poisson} are often contrasted with low-complexity \ac{BP}-based approaches, which scale linearly with the number of targets and measurements~\cite{meyer}. This motivates combining \ac{BP} with target handover. 

In this paper, we propose a novel \ac{BP}-based handover mechanism for \ac{MTT} in \ac{DISAC} systems. 
Our specific contributions are as follows: \textit{(i)} we introduce a new handover algorithm for \ac{BP}-based multi-cell \ac{MTT}, incorporating a principled message-passing criterion based on target belief states; \textit{(ii)} we formulate the handover process using factor graphs, enabling compatibility with existing \ac{BP}-\ac{MTT} approaches while minimizing inter-\ac{BS} communication; \textit{(iii)} we systematically compare centralized, distributed, and handover tracking architectures, demonstrating that our approach matches the accuracy of centralized tracking at a fraction of the communication cost; and \textit{(iv)} we release an open-source simulation framework to facilitate reproducibility and further exploration in the community. It is located at \text{https://github.com/BaiLiping/BPTargetHandover}.

\section{System Model}
 In this section, we describe the kinematic state of the targets and the measurement model. 

\subsubsection*{Target State Model} At discrete time \( k \), each target \( i \) is represented by a kinematic state vector \(\mathbf{x}_k^{i}\) that comprises its position and velocity. The state transition density is given by \(f(\mathbf{x}_{k}^i|\mathbf{x}_{k-1}^i)\). 
In this paper, we assume the constant velocity and stochastic acceleration kinematic model \cite{bar2004estimation}, \(\mathbf{x}_{k}^{i} = \mathbf{F} \mathbf{x}_{k-1}^i + \mathbf{v}_{k-1}^i\), where  \( \mathbf{F} \) is the state transition matrix and \( \mathbf{v}_k^i \) the additive zero-mean Gaussian process noise with covariance matrix \( \mathbf{Q} \).

\subsubsection*{Measurement Model} In the  \ac{DISAC} system, there are \( n_s>1 \) \acp{BS} with known positions and partially overlapping \acp{FoV}. At time \( k \), a set with \(n_m\) measurements 
\(
\mathbf{Z}_{k,s} = [\, \mathbf{z}_{k,s}^1, \ldots, \mathbf{z}_{k,s}^{n_m} \,],
\)
is obtained by the \ac{BS}~\( s \). Each element in the set can originate from a target or clutter, where clutter is modeled by \ac{PPP} with intensity \( c(\mathbf{z})\).
We introduce a detection probability function on the kinematic state \( p_d^s(\mathbf{x}_k^{i}) \in [0,1]\). To model the \ac{FoV} of \ac{BS}~\( s \),  we specify the following: if \(\mathbf{x}_k^{i} \in \mathrm{FoV}_s\), then $p_d^s(\mathbf{x}_k^{i})>0$, and $p_d^s(\mathbf{x}_k^{i})=0$ otherwise. 
When  a measurement \( \mathbf{z}_{k,s}^{m} \) originates from target \( i \), the measurement \( \mathbf{z}_{k,s}^{m} \) 
is modeled as 
\(
\mathbf{z}_{k,s}^{m} = \mathbf{h}_s\left(\mathbf{x}_k^i\right) + \mathbf{w}_{k,s}^i,
\)
where \( \mathbf{h}_s(\cdot) \) is a nonlinear measurement function for \ac{BS} \( s \), and \( \mathbf{w}_{k,s}^i \) is the additive zero mean Gaussian measurement noise with covariance matrix \(\mathbf{R} \). Measurement includes tuples comprising the \ac{ToA} and \ac{AoA} of channel paths \cite{venugopal2017channel}, obtained through standard channel parameter estimation methods \cite{jiang2021beamspace}.  
The equivalent likelihood function for measurement originating from a known target is:  $f(\mathbf{z}_{k,s}^m | \mathbf{x}_k^i) = \mathcal{N}(\mathbf{z}_{k,s}^m; \mathbf{h}_s(\mathbf{x}_k^i), \mathbf{R})$.

\section{\ac{BP}-Based Multi-Sensor Multi-Target Tracking}
\label{bp}

To introduce \ac{BP}-based \ac{MTT} in a \ac{DISAC} setup, we first briefly introduce factor graphs and the sum-of-product algorithm. Then, we discuss the case of a single-base-station \ac{BP}-\ac{MTT}. Finally, we explain centralized and distributed processing in the context of multiple-base-station \ac{BP}-\ac{MTT}.

\subsection{Factor Graph and \ac{BP} Basics}
Factor graphs convey the computational structure of a joint \ac{pdf} $f(x_1,\ldots,x_N)$ and are widely used for marginal posterior computations in robotics, communication, sensing, and machine learning~\cite{910572}. In a factor graph, factors (usually depicted as squares) and variables (traditionally shown as circles) are connected when a function takes that variable as input. \ac{BP} is a message passing method on a factor graph, whereby messages (functions) are computed and passed between factors and variables, which provides an efficient way to determine the marginals (also called the beliefs) $f(x_i)$, $i\in\{1,\ldots,N\}$.

\subsection{Single \ac{BS} \ac{MTT}} \label{sec:singleBS}
We first consider the case of a single \ac{BS}. We will introduce an augmented state to capture the unknown number of targets. Then we detail the prior, the likelihood, the factorization of the joint density, and the beliefs, largely based on \cite[Sections IV and VIII]{meyer}. Our focus will be on one time step (from \(k-1\) to \(k\)), to simplify the notation, at the cost of some mathematical rigour. 

\subsubsection{Augmented States}
To model the unknown and time-varying number of targets, we augment the kinematic state \( \mathbf{x}_k^j\) with an auxiliary binary variable \(r_k^j\), where \(r_k^j=1\) indicates the existence of the \ac{PT} and \(r_k^j=0\) otherwise. The augmented state of a \ac{PT} is denoted by \( \mathbf{y}_k^j=[\mathbf{x}_k^j,r_k^j]\).
We distinguish between \textit{new \acp{PT}} and \textit{legacy \acp{PT}}, where every measurement is treated as an initial detection of a new \ac{PT}, and these new \acp{PT} become legacy \acp{PT} in subsequent time steps. 
We denote the augmented state for new \ac{PT} by \(\mathbf{\overline{y}}_k\), and the augmented state for legacy \ac{PT} by \(\mathbf{\underline{y}}_k\). We denote the vector for all \acp{PT} by \(\mathbf{Y}_k\), with 
\(\mathbf{Y}_k=[\mathbf{\underline{Y}}_k, \overline{\mathbf{Y}}_k]\).

\subsubsection{Prediction and Prior}
We denote the posterior (or the belief) of the \(j\)th \ac{PT} at time \(k-1\) by \(f(\mathbf{\underline{y}}^j_{k-1})\). Each of the \acp{PT} at time \(k-1\) corresponds to a legacy \ac{PT} at time \(k\). The state transition density for the \(j\)th legacy \ac{PT} is
\begin{align}
& f(\underline{\mathbf{y}}_{k}^{j} | \underline{\mathbf{y}}_{k-1}^{j})
  = \begin{cases}     f_d({\mathbf{\underline{x}}}_{k}^{j})  & \underline{r}_{k}^{j}=0, \underline{r}_{k-1}^{j}=0,\\
     0 & \underline{r}_{k}^{j}=1, \underline{r}_{k-1}^{j}=0, \\ 
  (1 - p_s({\mathbf{\underline{x}}}_{k-1}^{j}))\, f_d(\mathbf{\underline{x}}_{k}^{j}) & \underline{r}_{k}^{j}=0, \underline{r}_{k-1}^{j}=1,\\
      p_s(\mathbf{\underline{x}}_{k-1}^{j})\, f(\underline{\mathbf{x}}_{k}^{j} | \mathbf{\underline{x}}_{k-1}^{j}) & \underline{r}_{k}^{j}=1, \underline{r}_{k-1}^{j}=1,
 \end{cases}
\end{align}

where \(p_s(\mathbf{\underline{x}}_{k-1}^{j})\) is the survival probability of the \ac{PT} and \(f_d(\underline{\mathbf{x}}_{k}^{j})\) is a dummy distribution (in the sense of being an arbitrary placeholder, that will be marginalized out when computing the beliefs). From the posterior of previous step  \(f(\mathbf{\underline{y}}^j_{k-1})\) and the transition density \(f(\underline{\mathbf{y}}_{k}^{j} | \mathbf{\underline{y}}_{k-1}^{j})\), we can derive the prior \ac{pdf} for the \(j\)th legacy \ac{PT} at time \(k\) as 
\(
f(\mathbf{\underline{y}}_k^{j})= 
\int f(  \mathbf{\underline{y}}_k^{j} | \mathbf{{\underline{y}}}_{k-1}^{j}){f}( \mathbf{\underline{y}}_{k-1}^{j}) \mathrm{d} {\mathbf{\underline{y}}_{k-1}^{j}}\).

The prior for the \(m\)th new \ac{PT} is set to be \(f_n(\overline{\mathbf{x}}_{k}^{m})\), which in this paper is heuristically determined from the \(m\)th measurement \(\mathbf{z}_k^m\).
\subsubsection{Likelihood}
To account for the unknown association between measurements and targets, we introduce two sets of discrete association variables:  track-oriented association variables \(\mathbf{a}_k\) and measurement-oriented association variables \(\mathbf{b}_k\). Specifically, \(a_k^j = m\) indicates that the \(j\)th \ac{PT} is associated with measurement \(m\) (while \(a_k^j = 0\) indicates that the \(j\)th \ac{PT} is not detected), and \(b_k^m = j\) indicates that measurement \(m\) is associated with \(j\)th \ac{PT} (while \(b_k^m = 0\) indicates that measurement \(m\) is associated with the new \ac{PT}).
To ensure that \(\mathbf{a}_k\) and \(\mathbf{b}_k\) are mutually consistent, there is a constraint \(\Psi(\mathbf{a}_k,\mathbf{b}_k) \in \{0,1\}\). The binary compatibility function \(\Psi(\cdot)\) can be factorized \cite[Eq.~(21)]{meyer}, which makes it computationally efficient in the factor graph representation and the associated \ac{BP}. 

We now introduce two so-called \textit{pseudo-likelihood} functions for each legacy \ac{PT} (denoted by \(q(\cdot)\) for the \(j\)th legacy \ac{PT}) and for each new \ac{PT} (denoted by \(v(\cdot)\) for the  \(m\)th new \ac{PT}). 

 These are defined as 
\begin{align}
q(\underline{\mathbf{x}}_{k}^{j},\underline{r}_k^j, a_{k}^{j}; \mathbf{Z}_{k}) = 
\begin{cases}
    \frac{p_d(\underline{\mathbf{x}}_{k}^{j})}{{c}(\mathbf{z}_k^m)} f(\mathbf{z}_{k}^{m} | \underline{\mathbf{x}}_{k}^{j})  & a_{k}^{j} = m \neq 0, r_k^j=1,\\
    1 - p_d(\underline{\mathbf{x}}_{k}^{j})  & a_{k}^{j} = 0, r_k^j=1,  \\
    0 & a_{k}^{j} \neq 0, r_k^j=0, \\
    1 & a_k^j = 0, r_k^j=0.
\end{cases}
\end{align}
and
\begin{align}
    & v(\overline{\mathbf{x}}_{k}^{m},r_k^m, b_{k}^{m}; \mathbf{z}_{k}^{m}) \\
    & = 
    \begin{cases} 0 &  b_{k}^{m} \neq 0, r_k^j=1, \\ 
    \frac{\mu_n}{{c}(\mathbf{z}_k^m)} f_n(\overline{\mathbf{x}}_{k}^{m}) f(\mathbf{z}_{k}^{m} | \overline{\mathbf{x}}_{k}^{m}) &  b_{k}^{m} = 0, r_k^j=1,\\
    f_d(\overline{\mathbf{x}}_{k}^{m}) & r_k^j = 0.
\end{cases}\notag 
\end{align}
where $\mu_n\ge 0$ is the mean number of newborn targets, which is Poisson distributed. The clutter intensity \(c(\mathbf{z})\) is the expected number of clutter in an infinitesimal region around \(\mathbf{z}\), and \(c(\mathbf{z}) = \mu_c\,f_c(\mathbf{z})\), where \(f_c(\mathbf{z})\) is the clutter \ac{pdf}.

\subsubsection{Factorization}

With the pseudo-likelihoods defined, we can factorize the joint \ac{pdf} at time step \(k\) based on conditional indepence as 
\begin{align}
& f(\mathbf{Y}_k, \mathbf{a}_k, \mathbf{b}_k, \mathbf{Z}_k) \\
& \propto  \Psi(\mathbf{a}_k,\mathbf{b}_k)\prod_{j=1}^{n_p} f(\mathbf{\underline{y}}_k^{j})q(\underline{\mathbf{y}}_k^j,a_k^j;\mathbf{Z}_k)\prod_{m=1}^{n_m}v(\overline{\mathbf{y}}_k^m,b_k^m;\mathbf{z}_k^m), \notag
\end{align}

\begin{figure*}
    \centering
    \includegraphics[width=0.75\linewidth]{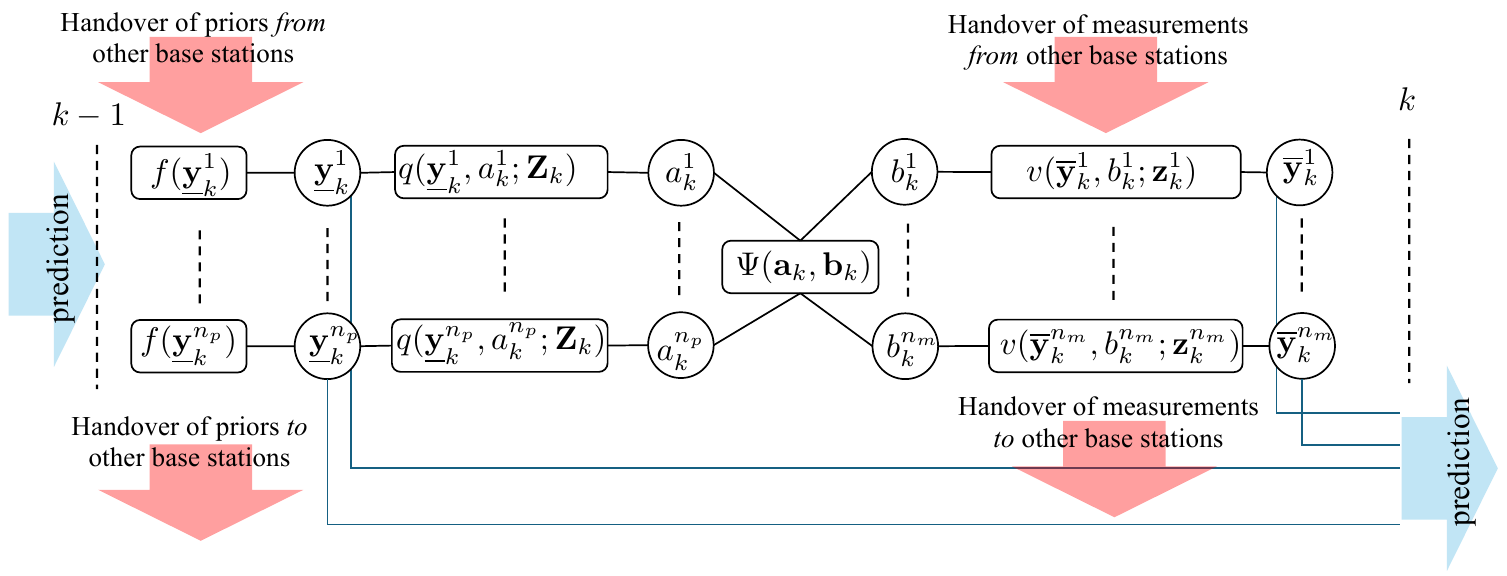}
    \caption{Factor graph representation of  $f(\mathbf{Y}_k, \mathbf{a}_k, \mathbf{b}_k, \mathbf{Z}_k)$. The prediction step can also be expressed as part of the factor graph, but is not shown for simplicity of the exposition. Handover of priors and measurements is shown in red.}
    \label{fig:BP}
    \vspace{-5mm}
\end{figure*}

The corresponding factorization is presented by the factor graph in Fig.~\ref{fig:BP}. 
The Bayesian estimation process is expressed in the messages passed in the factor graph from Fig.~\ref{fig:BP}. 
All the messages passed from the factor node to the variable node \(\mathbf{y}_k^j\) will be a weighted combination of all the cases (associations). We refer the readers to \cite{meyer} for the equations for each message.

\subsubsection{Posterior and Beliefs}

At time step \(k\), the inputs to the factor graph are the beliefs from time \(k-1\), as shown in Fig.~\ref{fig:BP}. For the \(j\)th legacy \ac{PT}, propagating its input beliefs through the loopy factor graph results in an approximated posterior density \(f(\underline{\mathbf{y}}_k^j | \mathbf{Z}_k)\). From this approximated posterior, two updated beliefs at time \(k\) can be obtained: one over the target’s kinematic state $\tilde{p}(\underline{\mathbf{x}}_k^j ) = \sum_{\underline{r}^j_k}f(\underline{\mathbf{y}}_k^j | \mathbf{Z}_k)$ and one over its existence $\tilde p(\underline{r}_k^j)=\int f(\underline{\mathbf{y}}_k^j | \mathbf{Z}_k) \mathrm{d}\underline{\mathbf{x}}_k^j$. Analogous beliefs are computed for each new \ac{PT}. 
The existence belief \(\tilde p(r_k^j=1 )\), also known as the posterior existence probability, is used to manage the \acp{PT}: we declare the \(j\)th \ac{PT} to exist if
\(
\tilde p(r_k^j=1 ) > P_{\mathrm{th}},
\)
and we prune that \ac{PT} if
\(
\tilde p(r_k^j=1 ) < P_{\mathrm{prune}}, 
\)
where \(P_\mathrm{th}\) and \(P_\mathrm{prune}\) are predetermined thresholds for declaring a target exists and pruning, respectively.

\subsection{Multi-\ac{BS} \ac{MTT}} 
\label{sequential}
Now, we extend our discussion from the single \ac{BS} to \ac{MTT} with \(n_s>1\) \acp{BP}. This is formally known as a multi-sensor \ac{MTT} problem. Each \ac{BS} \(s\) obtain a set with \(n_{m_s}\) measurements \(\mathbf{Z}_{k,s}\). There are now two extreme strategies:

\begin{itemize}
    \item \textit{Distributed Processing:} Each BS operates independently and executes the \ac{BP}-\ac{MTT} process outlined in Section \ref{sec:singleBS}. This means that when \acp{PT} leave the \ac{FoV} of a BS, they are lost to that BS in the sense that no more measurements will be associated with it. In addition, if that target enters the \ac{FoV} of another \ac{BS}, its track must be initialized from scratch.
    \item \textit{Centralized Processing:} In this case, there is a central unit (See Fig.~\ref{fig:overview}) to which all \acp{BS} send their measurements. There are several ways to proceed \cite{bar1995multitarget,vo_multi-sensor_2017, scalable-method-group}, but we present a simple sequential approach. In this sequential approach, the \acp{BS} are ordered and the central unit processes the measurements from each \acp{BS} sequentially, thereby generating new \acp{PT}, to which the measurements from the subsequent \acp{BS} can be associated. Effectively, sequentially updating the state beliefs in a sequence of  \(n_s\) \acp{BS} in one time is the equivalent of updating the state beliefs in \(n_s\) time in a distributed setup. 
\end{itemize}

\begin{example}[Fusion between two BSs]
To illustrate the sequential processing approach in the context of \ac{BP}-\ac{MTT}, we discuss the procedure for one new target located within the overlapping \ac{FoV} between \ac{BS}1 and \ac{BS}2. There are two measurements,  
$\mathbf{z}_{k,1}$ at BS1 and $\mathbf{z}_{k,2}$ at BS2,  originating from the target: \(\mathbf{z}_{k,1}\) is first used to initiate the new \ac{PT} with posterior \(f(\mathbf{\overline{y}}_k^1| \mathbf{z}_{k,1})\). Then \(\mathbf{z}_{k,2}\) is used to update the posterior \(f(\mathbf{\overline{y}}_k^1| \mathbf{z}_{k,1})\) to \(f(\mathbf{\underline{y}}_k^1 |\mathbf{z}_{k,1}, \mathbf{z}_{k,2})\). 

\end{example}

\section{\ac{BP}-Based Target Handover}
\label{sec:handover}
In this section, we introduce the proposed target handover mechanisms. By target handover, we mean sharing the target information from one BS to one or more BSs,  without deleting any information at transmission. The handover should not interfere with the \ac{BP}-\ac{MTT} process and avoid double-counting of measurements. 
We will present two target handover approaches: (i) with measurements exchange that exploits the overlapping \ac{FoV} of \acp{BS} to improve target state estimation while minimizing the information exchanged between \acp{BS} (illustrated in Fig.~\ref{fig:BP}); (ii)  without measurements exchange in order to highlight the benefits of sharing measurements in the numerical experiments. 

We focus on a single pair of \acp{BS} for notational clarity, though the methods extend to any configuration. We denote the \ac{BS} initiating the handover as \ac{BS}-Tx (index \(s_t\)) and the \ac{BS} receiving the information as \ac{BS}-Rx (index \(s_r\)). We assume parallel processing between the two \acp{BS}. That is, the handover priors are received by the \ac{BS}-Rx before prediction steps at time \(k\), and the handover measurements are received by the \ac{BS}-Tx before the updating steps at time \(k\).

\subsection{Operation at the Transmitting BS (\ac{BS}-Tx)}
From BS-Tx, the handover process requires keeping track of several additional variables, as described below:
\begin{enumerate}
    \item BS-Tx runs BP as described in Section \ref{sec:singleBS}. Hence, the target handover algorithm does not interrupt the \ac{BP}-\ac{MTT} process at \ac{BS}-Tx. 
    As a side-effect of running BP, we obtain the marginal association probabilities \(f^{(s_t)}(a_k^j|\mathbf{Z}_k)\). We can then obtain the index of the most likely associated measurement for the \(j\)th \ac{PT}, denoted \(m^j  = \arg \max_m f^{(s_t)}(a_k^j=m|\mathbf{Z}_k)\). If \(m^j=0\), the \ac{PT} is most likely to be miss-detected and no measurement is handed over. The notation $f^{(s_t)}$ indicates that this density is computed at BS $s_t$. 
    \item For each PT $j$, we check if the target handover criterion $h_c^j\in \{ 0,1\}$ is met:  
\begin{align}
& h_c^j = \mathbb{I}\{f^{(s_t)}(\underline r_{k}^j = 1)>P_{\mathrm{th}}\}\\
& \times 
\int
p_d^{(s_r)}(\underline{\mathbf x}_k^j)\,
f^{(s_t)}(\underline{\mathbf x}_k^j)\,
\mathrm{d}\underline{\mathbf x}_k^j > \Gamma, \notag
\end{align}

where  \(f^{(s_t)}(\underline{\mathbf x}_k^j)\) is the marginal of the prior \(f^{(s_t)}(\underline{\mathbf y}_k^j)\) over the kinematic state and \(f^{(s_t)}(\underline{r}_k^j)\) is the marginal of the prior over existence; \(\mathbb{I}\{\mathsf{P}\}\) is an indicator function valued \(1\) if $\mathsf{P}$ is true and \(0\) otherwise;  \(p_d^{(s_r)}(\underline{\mathbf x}_k^j)\) is the  \ac{BS}-Rx detection probability; \(\Gamma\) a predefined handover threshold.
\item \acp{PT} for which $h^j_c=1$ and $m^j>0$ are potential candidates for handover. To avoid repeated handovers of the same \ac{PT}, each \ac{PT} \(j\) is assigned a unique local label, say \(L^{(s_t)}_j\in \mathbb{N}\),  at \ac{BS}-Tx. Once a target has been handed over for the first time, its label is appended to a local list \(\mathcal{T}^{(s_t \to s_r)}\), where the subscript \((s_t \to s_r)\) indicates it is a list for handover labels from \ac{BS}-Tx to \ac{BS}-Rx. 
\item The actual handover now proceeds. If $h^j_c=1$ and \(L^{(s_t)}_j \notin \mathcal{T}^{(s_t \to s_r)}\), this is the first time we hand over this target from BS-Tx to BS-Rx. We hand over the prior \(f^{(s_t)}(\underline{\mathbf y}_k^j)\) and append \(L^{(s_t)}_j\) to \(\mathcal{T}^{(s_t \to s_r)}\). If \(m^j \neq 0\), we also hand over \(\mathbf{z}_k^{m_j}\) from \ac{BS}-Tx to \ac{BS}-Rx. On the other hand, if  $h^j_c=1$ and \(L^{(s_t)}_j \in \mathcal{T}^{(s_t \to s_r)}\), then this target was handed over before and we should not hand over the prior, to avoid double counting of information. Only the measurement $\mathbf{z}_k^{m_j}$ is handed over, provided $m^j\neq0$.
\item When a \ac{PT} is pruned or leaves the \ac{FoV} of \ac{BS}-Tx, its label is removed from \(\mathcal{T}^{(s_t \to s_r)}\).
\end{enumerate}

The target handover variant without measurements follows the same steps, but never hands over the measurements. 

\subsection{Operation at the Receiving BS (\ac{BS}-Rx)}
Now we describe the target handover algorithm for \ac{BS}-Rx.  At time \(k\), \ac{BS}-Rx receives from \ac{BS}-Tx a number of priors of \acp{PT} and a number of measurements. 
BS-Rx now incorporates this information as follows:
\begin{enumerate}
    \item The received priors are appended to the local priors, effectively increasing the number of legacy \acp{PT} in the factor graph. This leads to an increased complexity due to the larger number of legacy \acp{PT}, but allows BS-Rx to quickly detect new \acp{PT} entering its \ac{FoV}. 
     Each received \acp{PT} is associated with a unique local label \(L^{(s_r)}\), which is added to the local list \(\mathcal{R}^{(s_r \leftarrow s_t)}\) to prevent duplicate handovers.  
    \item BS-Rx runs \ac{BP} sequentially  (as detailed in Section \ref{sequential}), first its local measurements and then using the handed over received measurements from other BSs. 
    \item When a \ac{PT} is pruned or leaves the \ac{FoV} of \ac{BS}-Rx, its label is removed from \(\mathcal{R}^{(s_r \leftarrow s_t)}\). 
\end{enumerate}

The target handover without measurements for \ac{BS}-Rx follows the same steps, omitting only the measurement-related sequential processing.

\begin{rem}
    All \acp{BS} are acting as BS-Tx and BS-Rx at each time step and maintain transmission and reception lists with respect to all neighboring \acp{BS} and also share the information related to the \ac{FoV}, in the form of the spatial detection probability. 
    It is also important to highlight that priors are handed over before BP update at time step $k$, while measurements are handed over after BP update at time step $k$.
\end{rem}

\section{Numerical  Results}

We evaluate the proposed BP-based handover approach in a simulated \ac{DISAC} scenario and compare to distributed and centralized baselines. 

\subsection{Scenario}
\begin{figure}[t]
  \centering
  \includegraphics[width=0.8\linewidth]{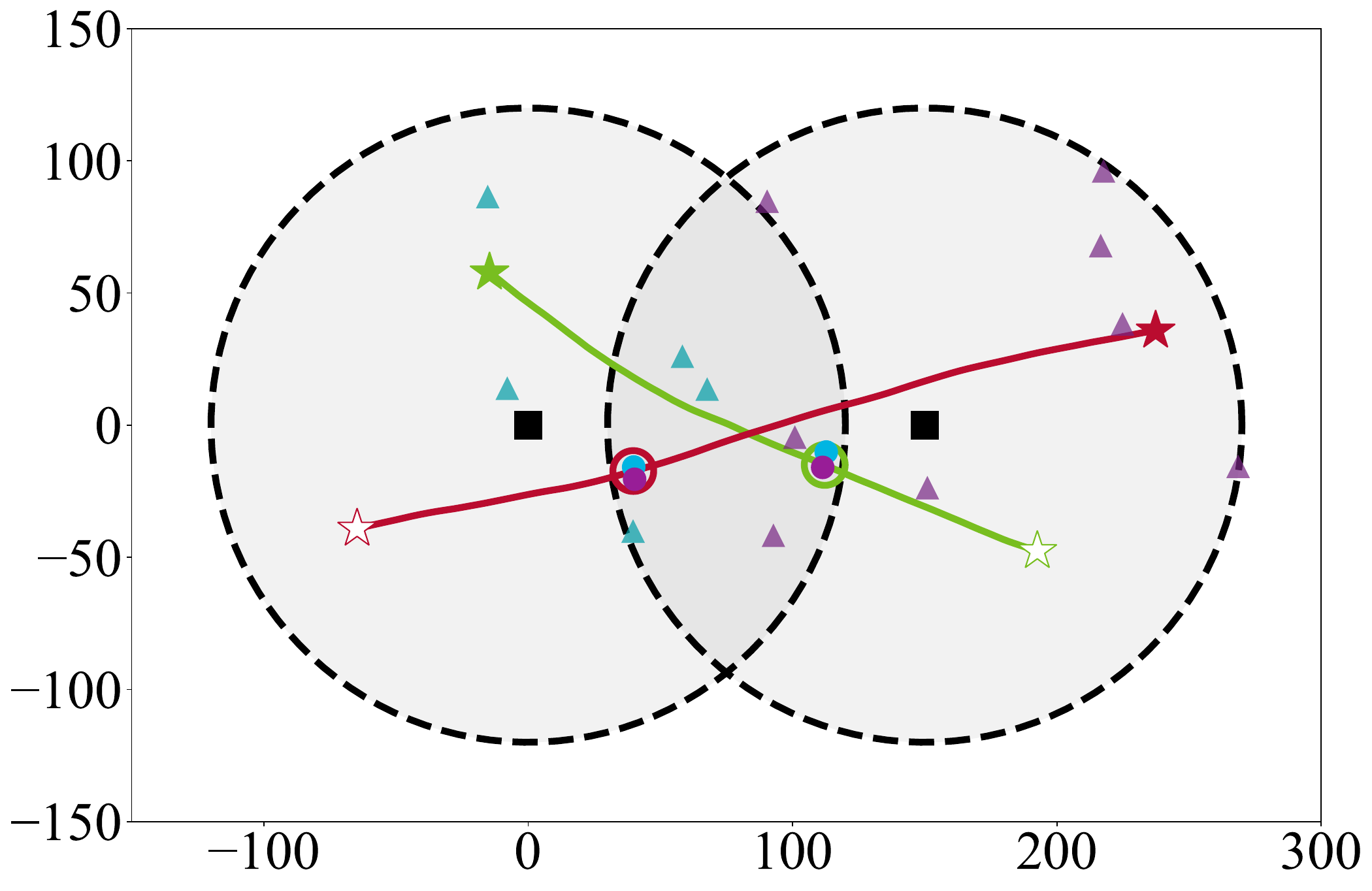}
  \scalebox{0.6}{
  \begin{tikzpicture}[font=\footnotesize\rmfamily]
    \node[inner sep=0] {
      \begin{tabular}{@{} ll @{\quad} ll @{\quad} ll @{}}
        \tikz{\filldraw[black] (0,0) rectangle (1ex,1ex);} & base station &
        \tikz{\draw[black,dashed,line width=1pt] (0.5ex,0.5ex) circle (1ex);} & FoV &
        \tikz{\node[star,star points=5,star point ratio=2,draw=black,fill=black,inner sep=0,minimum size=2ex]{};} & start position \\

        \tikz{\node[star,star points=5,star point ratio=2,draw=black,fill=white,inner sep=0,minimum size=2ex]{};} & end position &
        \tikz{\draw[line width=1.5pt, draw=green!50!black] (0,0) -- ++(1.5ex,0);} & target 1 trajectory &
        \tikz{\draw[line width=1.5pt, draw=red] (0,0) -- ++(1.5ex,0);} & target 2 trajectory \\

        \tikz{\draw[line width=1pt, draw=green!50!black, fill=white] (0,0) circle (1ex);} & target 1 current pos &
        \tikz{\draw[line width=1pt, draw=red, fill=white] (0,0) circle (1ex);} & target 2 current pos &
        \tikz{\node[circle, draw=none, fill=cyan, inner sep=0, minimum size=2ex]{};} & true meas. (BS1) \\

        \tikz{\node[regular polygon,regular polygon sides=3,draw=none,fill=teal,inner sep=0,minimum size=2ex]{};} & clutter (BS1) &
        \tikz{\node[circle, draw=none, fill=purple, inner sep=0, minimum size=2ex]{};} & true meas. (BS2) &
        \tikz{\node[regular polygon,regular polygon sides=3,draw=none,fill=violet,inner sep=0,minimum size=2ex]{};} & clutter (BS2) \\
      \end{tabular}
    };
  \end{tikzpicture}}

  \caption{Simulation scenario with 2 \acp{BS} and two targets. Axes are in meters. }
  \label{simulation}
\end{figure}
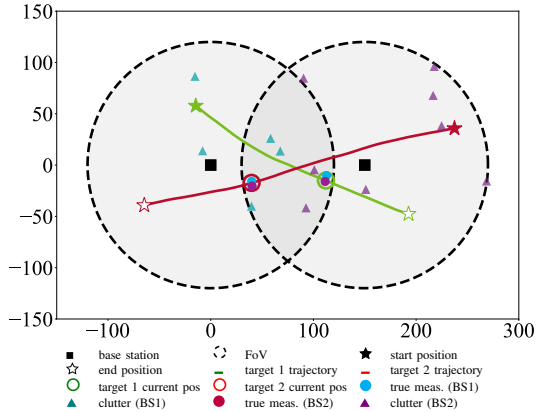

Two \acp{BS}, denoted BS1 and BS2, are located at \([0\mathrm{m},0\mathrm{m}]\) and \([150\mathrm{m},0\mathrm{m}]\), respectively. Each \ac{BS} has a circular \ac{FoV} with radius \(120\)\,m centered at its position. Two moving targets traverse the combined surveillance region; their start and end positions, as well as their trajectories, are shown in Fig.~\ref{simulation}. Simulations considered a sampling interval of \( 1\,\mathrm{s}\) over \(100\) time steps. From about \(35\ \mathrm{s}\) to \(70\ \mathrm{s}\), there are on average two targets. The target handover is initiated around \(25\ \mathrm{s}\) to \(35\ \mathrm{s}\). Between time steps \(47\) and \(52\), the two targets pass close to one another, as illustrated in Fig.~\ref{simulation}.
For data generation, we use the following settings for both \acp{BS}: \(p_d(\mathbf{x}) = 1\) when \( \mathbf{x} \in \mathrm{FoV}\) and \(p_d(\mathbf{x}) = 0\) otherwise; a constant survival probability \(p_s(\mathbf{x})=0.99\); to set $\mathbf{Q}$, we use the model  \cite[eq.~6.3.2-4]{bar2004estimation} with process noise standard deviation \( \sigma_v = 0.05~\mathrm{m}\) on both \(x\) and \(y\) coordinates; measurement noise covariance $\mathbf{R}=\text{diag}(\sigma^2_r,\sigma^2_{\theta})$  where  \( \sigma_r = 1~\mathrm{m}\) and \( \sigma_{\theta} = 1^\circ\) for \ac{ToA} (converted to distance) and \ac{AoA}, respectively. The value of \( \mu_c\) is set to be \(5\) and the clutter is distributed uniformly over the respective \ac{FoV}.

\subsection{\ac{BP}-\ac{MTT} Implementation}
We implemented the \ac{BP}-\ac{MTT} methods in both particle filter and \ac{EKF}, and they can be found in the provided open source code. We will discuss the particle filter and display its results. All the parameters are set to be the same for both \acp{BS}. For the particle filter implementation, the number of particles is $10,000$. The threshold for pruning is \(P_\mathrm{prune}=10^{-5}\), and the detection threshold is set to \(P_\mathrm{th}=0.5\). The handover criterion \(\Gamma\) is \(0.5\). The \(\mu_n\) is set to be \(0.01\), and the new target is uniformly distributed in the \ac{FoV}. Different from the generative model, we set \(p_d(\mathbf{x}) = 0.9\) when \( \mathbf{x}\) is within the \acp{FoV} and \(p_d(\mathbf{x}) = 0\) otherwise. In addition, \(p_s(\mathbf{x})\), \( \sigma_{\theta}\), \( \sigma_{r}\), \( \sigma_{r}\), \( \mu_c\) are all set to be identical to that of the data generation process. 

We implemented four algorithms for comparison:  distributed processing (from Section \ref{sec:singleBS}), centralized processing (from Section \ref{sequential}), and the proposed target handover with and without measurements (from Section \ref{sec:handover}).  For each of the four algorithms, we ran 100 Monte Carlo (MC) trials and evaluated performance via the \ac{GOSPA} \cite{GOSPA} metric.\footnote{The overall \ac{GOSPA} score consists of three components: missed-detection score (no detection can be associated with the true track), false alarm score (detection when there is no track), and the localization score (Euclidean distance between the true track and the closest assigned estimation). The lower the overall \ac{GOSPA} score, the better the performance of the proposed methods. The detailed GOSPA implementation and parameters are provided in the simulation code.}

\begin{figure}[t]
  \centering
  \begin{tikzpicture}
    \begin{axis}[
      name=pltA,
      width=\linewidth, height=5cm,
      xlabel={time (s)},
      ylabel={GOSPA score},
      grid=both,
      tick label style={font=\footnotesize\rmfamily},
      label style={font=\footnotesize\rmfamily},
      title style={font=\footnotesize\rmfamily},
      legend style={
        at={(axis description cs:0.98,0.98)},
        anchor=north east,
        legend columns=1,
        font=\footnotesize\rmfamily
      }
    ]
      \addplot[thick,color=blue]           table[x=Time,y=Centralized_gospa,col sep=comma] {BS1_subplots/bs1_combined.csv};
      \addplot[thick,color=green!70!black] table[x=Time,y=Distributed_gospa,col sep=comma] {BS1_subplots/bs1_combined.csv};
      \addplot[thick,color=orange]         table[x=Time,y=HandoverNoMeas_gospa,col sep=comma] {BS1_subplots/bs1_combined.csv};
      \addplot[thick,color=red]            table[x=Time,y=HandoverMeas_gospa,col sep=comma] {BS1_subplots/bs1_combined.csv};
      \addlegendentry{centralized}
      \addlegendentry{distributed}
      \addlegendentry{handover w/o Meas}
      \addlegendentry{handover w/ Meas}
      \addlegendimage{dashed,thick,gray}
      \addlegendentry{avg. targets}
    \end{axis}
    \begin{axis}[
      at={(pltA.north east)}, anchor=north east,
      width=\linewidth, height=5cm,
      axis x line=none,
      axis y line*=right,
      ylabel={avg.\ targets},
      ytick={1,2},
      ytick pos=right,
      tick label style={font=\footnotesize\rmfamily},
      label style={font=\footnotesize\rmfamily},
      y label style={xshift=-5pt},
    ]
      \addplot[dashed,thick,gray]
        table[x=Time,y=AvgTargets,col sep=comma]
        {BS1_subplots/bs1_targets.csv};
    \end{axis}
  \end{tikzpicture}
  \caption{GOSPA score for \ac{BS}1.}
  \label{fig:gospa}
\end{figure}
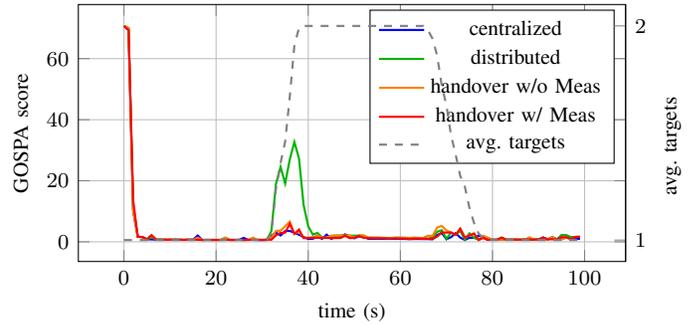

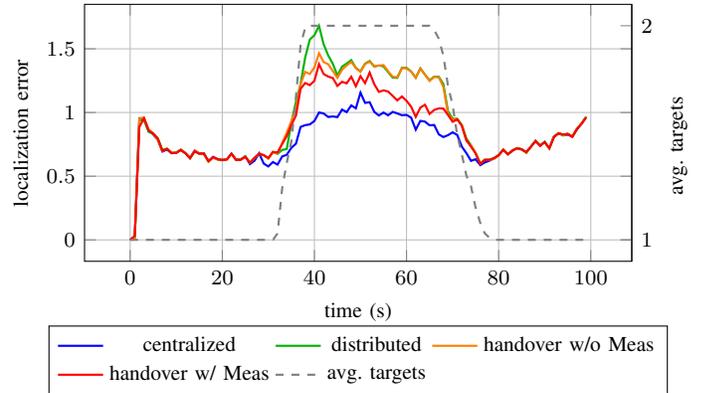
\begin{figure}[t]
  \centering
  \begin{tikzpicture}
    \begin{axis}[
      name=pltB,
      width=\linewidth, height=5cm,
      xlabel={time (s)},
      ylabel={localization error},
      grid=both,
      tick label style={font=\footnotesize\rmfamily},
      label style={font=\footnotesize\rmfamily},
      title style={font=\footnotesize\rmfamily},
      legend style={
        at={(0.5,-0.25)}, anchor=north,
        legend columns=3,
        font=\footnotesize\rmfamily
      }
    ]
      \addplot[thick,color=blue]           table[x=Time,y=Centralized_localization,col sep=comma] {BS1_subplots/bs1_combined.csv};
      \addplot[thick,color=green!70!black] table[x=Time,y=Distributed_localization,col sep=comma] {BS1_subplots/bs1_combined.csv};
      \addplot[thick,color=orange]         table[x=Time,y=HandoverNoMeas_localization,col sep=comma] {BS1_subplots/bs1_combined.csv};
      \addplot[thick,color=red]            table[x=Time,y=HandoverMeas_localization,col sep=comma] {BS1_subplots/bs1_combined.csv};
      \addlegendentry{centralized}
      \addlegendentry{distributed}
      \addlegendentry{handover w/o Meas}
      \addlegendentry{handover w/ Meas}
      \addlegendimage{dashed,thick,gray}
      \addlegendentry{avg. targets}
    \end{axis}
    \begin{axis}[
      at={(pltB.north east)}, anchor=north east,
      width=\linewidth, height=5cm,
      axis x line=none,
      axis y line*=right,
      ylabel={avg.\ targets},
      ytick={1,2},
      ytick pos=right,
      tick label style={font=\footnotesize\rmfamily},
      label style={font=\footnotesize\rmfamily},
      y label style={xshift=-5pt},
    ]
      \addplot[dashed,thick,gray]
        table[x=Time,y=AvgTargets,col sep=comma]
        {BS1_subplots/bs1_targets.csv};
    \end{axis}
  \end{tikzpicture}
  \caption{Localization error for BS1.}
  \label{fig:loc}
\end{figure}

\begin{figure}[t]
  \centering
  \begin{tikzpicture}
    \begin{axis}[
      name=pltC,
      width=\linewidth, height=5cm,
      xlabel={time (s)},
      ylabel={missed-detection error},
      grid=both,
      tick label style={font=\footnotesize\rmfamily},
      label style={font=\footnotesize\rmfamily},
      title style={font=\footnotesize\rmfamily},
      legend style={
        at={(axis description cs:0.98,0.98)},
        anchor=north east,
        legend columns=1,
        font=\footnotesize\rmfamily,
      }
    ]
      \addplot[thick,color=blue]           table[x=Time,y=Centralized_miss_truth,col sep=comma] {BS1_subplots/bs1_combined.csv};
      \addplot[thick,color=green!70!black] table[x=Time,y=Distributed_miss_truth,col sep=comma] {BS1_subplots/bs1_combined.csv};
      \addplot[thick,color=orange]         table[x=Time,y=HandoverNoMeas_miss_truth,col sep=comma] {BS1_subplots/bs1_combined.csv};
      \addplot[thick,color=red]            table[x=Time,y=HandoverMeas_miss_truth,col sep=comma] {BS1_subplots/bs1_combined.csv};
      \addlegendentry{centralized}
      \addlegendentry{distributed}
      \addlegendentry{handover w/o Meas}
      \addlegendentry{handover w/ Meas}
      \addlegendimage{dashed,thick,gray}
      \addlegendentry{avg. targets}
    \end{axis}
    \begin{axis}[
      at={(pltC.north east)}, anchor=north east,
      width=\linewidth, height=5cm,
      axis x line=none,
      axis y line*=right,
      ylabel={avg.\ targets},
      ytick={1,2},
      ytick pos=right,
      tick label style={font=\footnotesize\rmfamily},
      label style={font=\footnotesize\rmfamily},
      y label style={xshift=-5pt},
    ]
      \addplot[dashed,thick,gray]
        table[x=Time,y=AvgTargets,col sep=comma]
        {BS1_subplots/bs1_targets.csv};
    \end{axis}
  \end{tikzpicture}
  \caption{missed-detection error for \ac{BS}1.}
  \label{fig:missing}
\end{figure}
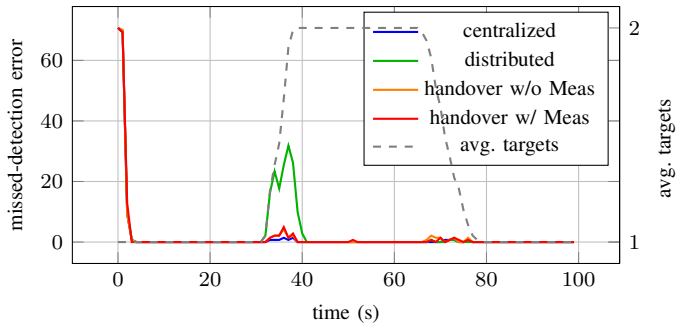

\subsection{Results and Discussion}

 The scores for both \ac{BS}1 and \ac{BS}2 are evaluated with estimates and true tracks within their respective \ac{FoV}. We present only the results for \ac{BS}1 (results for \ac{BS}2 are available in the code).

The overall \ac{GOSPA} score for \ac{BS}1 is shown in Fig.~\ref{fig:gospa}.
The proposed target handover with measurements algorithm achieves an overall \ac{GOSPA} score comparable to that of the centralized approach, thanks largely to its reduction in missed‐detection error. Since the missed‐detection error dominates the \ac{GOSPA} decomposition, the overall GOSPA curve closely mirrors the missed‐detection curve. In the following paragraphs, we will analyse each of the three contributors to the \ac{GOSPA} score—localization error, missed‐detection error, and false‐alarm error—in detail.

The localization error component in the GOSPA for \ac{BS}1, shown in Fig.~\ref{fig:loc}, demonstrates that within the overlapping \ac{FoV}, the centralized algorithm achieves the lowest error, followed by the target handover with measurements algorithm. The handover without measurements approach yields only a slight error reduction at the start of the handover before its performance converges to that of the fully distributed method. This underscores the benefit of exchanging measurements during handover. All algorithms exhibit increased localization errors in the overlap region, where the reduced separation between targets makes data association more challenging.

The missed-detection error component for \ac{BS}1 is shown in Fig.~\ref{fig:missing}. All algorithms exhibit elevated missed-detection errors in the initial frames, since it takes several steps for the posterior existence probability to exceed the detection threshold \(P_{\mathrm{th}}\). In the fully distributed case, a pronounced spike in missed detections occurs when the target first moves into the \ac{FoV} of \ac{BS}1, because a new track must be initiated locally. Both the target handover without measurements and the target handover with measurements schemes achieve missed-detection performance comparable to that of the centralized algorithm.

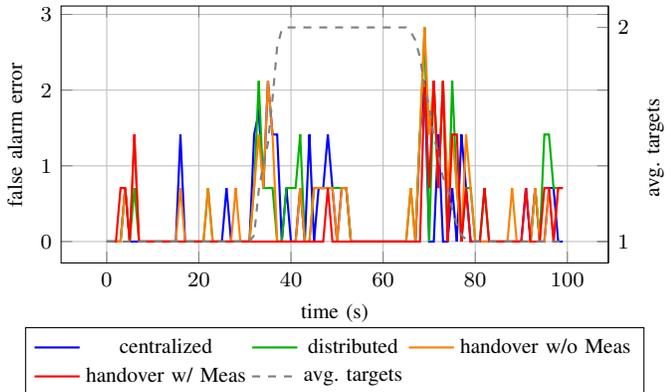
\begin{figure}
  \centering
  \begin{tikzpicture}
    \begin{axis}[
      name=pltD,
      width=\linewidth, height=5cm,
      xlabel={time (s)},
      ylabel={false alarm error},
      grid=both,
      tick label style={font=\footnotesize\rmfamily},
      label style={font=\footnotesize\rmfamily},
      title style={font=\footnotesize\rmfamily},
      legend style={
        at={(0.5,-0.25)}, anchor=north,
        legend columns=3,
        font=\footnotesize\rmfamily
      }
    ]
      \addplot[thick,color=blue]           table[x=Time,y=Centralized_false_tracks,col sep=comma] {BS1_subplots/bs1_combined.csv};
      \addplot[thick,color=green!70!black] table[x=Time,y=Distributed_false_tracks,col sep=comma] {BS1_subplots/bs1_combined.csv};
      \addplot[thick,color=orange]         table[x=Time,y=HandoverNoMeas_false_tracks,col sep=comma] {BS1_subplots/bs1_combined.csv};
      \addplot[thick,color=red]            table[x=Time,y=HandoverMeas_false_tracks,col sep=comma] {BS1_subplots/bs1_combined.csv};
      \addlegendentry{centralized}
      \addlegendentry{distributed}
      \addlegendentry{handover w/o Meas}
      \addlegendentry{handover w/ Meas}
      \addlegendimage{dashed,thick,gray}
      \addlegendentry{avg. targets}
    \end{axis}
    \begin{axis}[
      at={(pltD.north east)}, anchor=north east,
      width=\linewidth, height=5cm,
      axis x line=none,
      axis y line*=right,
      ylabel={avg.\ targets},
      ytick={1,2},
      ytick pos=right,
      tick label style={font=\footnotesize\rmfamily},
      label style={font=\footnotesize\rmfamily},
      y label style={xshift=-5pt},
    ]
      \addplot[dashed,thick,gray]
        table[x=Time,y=AvgTargets,col sep=comma]
        {BS1_subplots/bs1_targets.csv};
    \end{axis}
  \end{tikzpicture}
  \caption{False alarms error for BS1.}
  \label{fig:false}
\end{figure}

The false alarm error component for \ac{BS}1 is presented in Fig.~\ref{fig:false}. We observe elevated false alarm rates at the boundary of the overlapping \ac{FoV} for both the target handover with measurements and the target handover without measurements schemes. The timing of the target handover is determined by the handover threshold \(\Gamma\). Increasing \(\Gamma\) would likely decrease false alarm errors at the cost of higher missed-detection errors. Within the overlapping \ac{FoV}, the target handover with measurements approach achieves significantly fewer false alarms than the target handover without measurements case.

\section{Conclusions}

We have proposed a novel belief-propagation-based target handover procedure for \ac{MTT} in \ac{DISAC} systems. Our approach addresses the need for scalable, decentralized tracking in scenarios where targets move across overlapping fields of view covered by spatially distributed base stations. By leveraging factor graph representations and message-passing principles, the proposed method enables efficient target handover with limited communication overhead.
Through detailed simulations, we have shown that the proposed handover mechanism achieves tracking performance comparable to centralized processing, while avoiding its high bandwidth and computational requirements. The algorithm maintains track continuity across base stations and is well-suited for dense urban environments where scalability and low latency are critical.
Future work will focus on extending the handover procedure to support more complex and dynamic sensing configurations and real-time operation over heterogeneous networks. In addition, integrating this method into standardized communication frameworks for ISAC will be an important step toward practical deployment in 6G systems.

\bibliographystyle{IEEEtran}
\bibliography{reference}
\end{document}